\documentclass[prl,twocolumn,showpacs,preprintnumbers,amsmath,amssymb,floatfix]{revtex4}
\usepackage{hyperref}
\usepackage{epsfig}
\begin{document}

\title{
Jahn-Teller Spectral Fingerprint in Molecular Photoemission: C$_{60}$
}

\author{Nicola Manini}\email{nicola.manini@mi.infm.it}
\author{Paolo Gattari}
\affiliation{Dipartimento di Fisica, Universit\`a di Milano,
Via Celoria 16, 20133 Milano, Italy}
\affiliation{INFM, Unit\`a di Milano, Milano, Italy}
\author{Erio Tosatti}
\affiliation{International School for Advanced Studies (SISSA), Via Beirut
4, 34014 Trieste, Italy}
\affiliation{INFM Democritos National Simulation Center, and INFM, Unit\`a
Trieste, Italy}
\affiliation{International Centre for Theoretical Physics (ICTP), P.O. Box
586, 34014 Trieste, Italy}

\date{\today}

\begin{abstract}

%
The $h_u$ hole spectral intensity for C$_{60} \to$~C$_{60}^+$
molecular photoemission is
calculated at finite temperature by a parameter-free Lanczos
diagonalization of the electron-vibration Hamiltonian, including the full 8
$H_g$, 6 $G_g$, and 2 $A_g$ mode couplings.
%
%
The computed spectrum at 800~K is in striking agreement with gas-phase
data.  The energy separation of the first main shoulder from the main
photoemission peak, 230~meV in C$_{60}$, is shown to measure directly and
rather generally the strength of the final-state Jahn-Teller coupling.
\end{abstract}


\maketitle

\begin{figure}[b!]
\centerline{
\epsfig{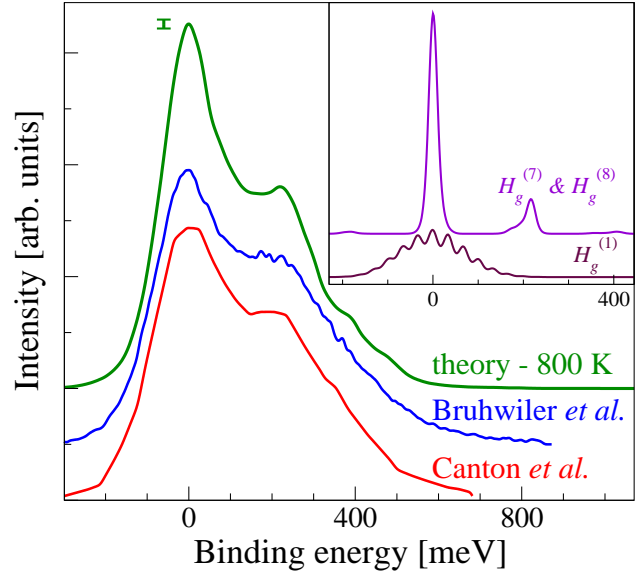}
}
\caption{\label{allmodesT800:fig}
Experimental PE spectra by Canton {\it et al.}~\cite{Canton02etal} and by
Br\"uhwiler {\it et al.}~\cite{Bruhwiler97etal} compared with the
calculation based on the {\it ab initio} parameters of
Ref.~\cite{Manini01etal} ($T=800$~K, $N_{\rm tier}=31$, $N_{\rm max}=10^5$,
$N_{\rm sample}=245$).
Intensities are normalized to unity.
%
%
Inset: $T=800$~K spectra including the single $H_g^{(1)}$ mode, and
$H_g^{(7)}$ and $H_g^{(8)}$.
All spectra are shifted to move the main peak to zero energy and
broadened with a 10~meV HWHM Gaussian.
The errorbar gives the largest statistical error introduced by random
sampling of the initial states.
}
\end{figure}

Photoemission (PE) from a closed-shell high-symmetry molecule, involving
Jahn-Teller (JT) effect in the final state, is accompanied by
characteristic vibronic structures in the measured hole spectrum
\cite{Lindner96etal}.
Although accurate spectral calculations have recently appeared in the
chemical literature, e.g.\ for benzene \cite{Koppel02etal}, there still is
a strong need for a qualitative understanding of a more general nature,
possibly system independent, and informative on the nature and the strength
of the JT coupled problem.
A recent case in point is that of gas-phase C$_{60}$, where an important
side peak was reported 230~meV above the main PE peak
\cite{Bruhwiler97etal,Canton02etal} (see Fig.~\ref{allmodesT800:fig}).
As this excitation energy is in fact larger than all vibrational
frequencies of C$_{60}$ (32-197~meV), its nature could not be
straightforwardly interpreted, leading to a
debate \cite{ManiniCommAndReply03}.

In this Letter we present first of all a detailed understanding of 
the PE structures that are accurately reproduced for
C$_{60} \to$~C$_{60}^+$ (Fig.~\ref{allmodesT800:fig}) 
through calculations that fully include temperature and that 
contain no adjustable parameters. 
By taking further
this calculation to pieces, we
identify the important ingredients that determine the spectrum.
The first is symmetry selection: the initial orbital symmetry dictated by 
the suddenly injected bare hole extends to the whole spectrum. 
The second is a property of weakly coupled 
JT vibronic 
multiplets that requires the first excitation
energy of the same symmetry as the ground state (GS) to rise above the bare
vibration energy by an amount  strictly
proportional to the squared JT coupling. 
Specifically in C$_{60}^+$, symmetry requires the 230~meV peak to
correspond to an $h_u$ vibronic excitation above the $h_u$ GS.
JT transforms the highest $H_g^{(8)}$ vibration ($\hbar\omega<200$~meV) of
C$_{60}$ into an $h_u$ vibron near 230~meV.
Interestingly, the 20\% magnitude ot this upward shift fingerprints quite
accurately the medium-sized overall JT coupling of C$_{60}$.
As both properties above are apparently general, and since the upward
shift tracks the coupling intensity well beyond the weak-coupling limit,
this kind of analysis appears of much more general value.
In particular, the first vibronic excitation energy as seen in PE of other
weak to medium JT-coupled molecules should and does gauge their respective
couplings as well.

%
%
We adopt for C$_{60}$ the simplest model \cite{CeulemansII,hbyh,Moate96etal} 
describing the JT coupling of the orbitally degenerate $h_u$ hole 
with the molecular vibrations.
The linear JT coupling is nonzero for two nondegenerate $A_g$, six
fourfold-degenerate $G_g$ and eight fivefold-degenerate $H_g$ vibration
modes, or 66 vibrations altogether \cite{CeulemansII,hbyh,Lueders02etal}. 
The resulting $h \otimes (A + G + H)$ JT Hamiltonian thus consists of 
the fivefold $h_u$ highest occupied molecular orbital (HOMO),
of 66 harmonic oscillators, and of the
electron-vibration (e-v) coupling term \cite{Manini01etal,hbyh}:
\begin{eqnarray}
\label{modelhamiltonian}
\hat{H} &=& \hat{H}_0 + \hat{H}_{\rm vib} + 
\hat{H}_{\rm e-v} \,,\\
\hat{H}_0     &=& \epsilon_{\rm HOMO} \, \sum_{m \sigma}  
\hat{c}_{m \sigma}^\dagger \hat{c}_{m \sigma} \,, \\
\label{vib-hamiltonian}
\hat{H}_{\rm vib} &=& \frac 12 \sum_{\Lambda j \mu} \hbar \omega_{\Lambda j}
\left(\hat{P}_{\Lambda j \mu}^2+\hat{Q}_{\Lambda j \mu}^2\right) \,, \\
\label{JT-hamiltonian}
\hat{H}_{\rm e-v} &=& 
\sum_{{r\,\Lambda j \mu}\atop{\sigma m m'}}
k^\Lambda g^{(r)}_{\Lambda j} \hbar \omega_{\Lambda j}
C^{r \Lambda \mu}_{m \; -m'} \,
\hat{Q}_{i\Lambda \mu}
\hat{c}^\dagger_{m\sigma} \hat{c}_{m' \sigma } \,.
\end{eqnarray}
%
%
Here $m$, $\mu$ label components within the degenerate multiplets
\cite{hbyh,Butler81},
$j$ counts modes of symmetry $\Lambda$,
$C^{r \Lambda \mu}_{m m'}$ are Clebsch-Gordan coefficients \cite{Butler81}
of the icosahedral group $I_h$ that couple the $h_u$ fermion
operators $\hat{c}_{m \sigma}^\dagger$ to a $\Lambda$ vibration.
$\hat{Q}_{i\Lambda \mu}
$
are the dimensionless 
normal-mode vibration coordinates
in units of 
$x_0(\omega_{\Lambda j})=\sqrt{\hbar/
(\omega_{\Lambda j} \, m_{\rm C})}$ 
where $m_{\rm C}$ is the mass of the C atom), and $\hat{P}_{i\Lambda \mu}$
the corresponding conjugate momenta.
The additional multiplicity $r=1,2$, needed for $H_g$ vibrations only,
labels the two separate kinds of $H_g$ coupling allowed under the {\em
same} symmetry \cite{Manini01etal,Butler81}.

In the present calculation we adopt the fully {\it ab initio} numerical
values of the e-v coupling parameters
$g^{(r)}_{\Lambda j}$
obtained by a Density Functional (DF) calculation in Ref.~\cite{Manini01etal}
\footnote{Another DF calculation \cite{Saito02} based on a different functional
yielded couplings roughly in the same range.
However we could not re-compute the PE spectrum with these couplings due to
their lack of distinction between coupling contributions of type $r=1$ and
$r=2$.}.
%
We also use the calculated vibrational frequencies of
Ref.~\cite{Manini01etal}, that are close to the experimental frequencies of
the $A_g$ and $H_g$ modes, but also cover the experimentally unavailable or
unreliable $G_g$ modes.  E-v couplings are dimensionless, each of them
normalized
to its respective vibration quantum, so as to reflect directly the
relative coupling strength.
%
%
Numerical factors 
$k^{A_g}=\frac {5^{\frac 12}}2$,
$k^{G_g}=\frac {5^{\frac 12}}4$,
$k^{H_g}=\frac 12$,
are included to make contact 
with 
Ref.~\cite{Manini01etal}.

In the sudden approximation, the angle-integrated PE spectrum $I(E)$ is
given by Fermi's golden rule \cite{Koppel02etal}
\begin{eqnarray}
\label{thermalaverage}
I(E) &=& \frac 1{10} \sum_{m\sigma i} I_{im\sigma}(\omega) P(i) \,,\\
\label{FermiGR}
I_{im\sigma}(E) &=& 
\sum_f \left|\langle f| \hat{c}_{m \sigma}^\dagger | i \rangle \right|^2
\delta\left(E - E_f+E_i\right) \,,
\end{eqnarray}
where $| i \rangle $ represents the starting vibrational eigenstate of
neutral C$_{60}$, of energy $E_i$, and $| f \rangle$ represents a vibronic
JT eigenstate of C$_{60}^+$, of energy $E_f$.
%
%
$\hat{c}_{m \sigma}^\dagger | i \rangle$ is the initial bare hole, where
the spin-$\sigma$ electron has been ejected from one orbital $m$ of the
fivefold degenerate $h_u$ HOMO, but
the molecule is still unrelaxed.
The factor 1/10 and the sum over $m$ and $\sigma$ signifies averaging over
the 5 orbital and 2 spin states of the hole.
Assuming thermal equilibrium we generate the
boson numbers $\{v_{\Lambda j\mu}\}$ of the initial neutral molecule
by randomly sampling the probability distribution
$
 P(i)= Z^{-1}\exp\left(-E_i / k_{\rm B}T\right) ,
$
where $Z=\sum_i\exp\left(-E_i / k_{\rm B}T\right)$ and
$E_i = \sum \hbar \omega_{\Lambda j} v_{\Lambda j\mu}$.
In the present medium-coupling regime we compute the final-state vibronic
energies and matrix elements in \eqref{FermiGR} by numerical Lanczos
diagonalization of the Hamiltonian \eqref{modelhamiltonian} on the product
basis of the five $h_u$ states times the harmonic oscillator ladders.
Since these contain an infinite number of states, some truncation is
necessary.
%
%
As we generally want to address high temperatures of the molecular beam
($T=800$~K used for C$_{60}$, $k_{\rm B} T\gtrsim 2\hbar\omega$ of the
strongly coupled $H_g^{(1)}$ mode)
a standard truncation would involve a far too large Hilbert space size.
We generate a smarter basis starting from the initial excitation
$\hat{c}_{m \sigma}^\dagger | i \rangle$, and iteratively
adding sets of states (``tiers'') directly coupled to those of the previous
tier by matrix elements of $\hat{H}_{\rm e-v}$, in a scheme inspired by
Ref.~\cite{Stuchebrukhov93}.
This procedure is iterated $N_{\rm tier}$ times so that states relevant at
up to $(2 N_{\rm tier})$-th order in perturbation theory are included
\footnote{
The tier size increases quickly, but we cut it off by including up
to some given number $N_{\rm max}$ of states per tier, according to their
perturbative weight $w_b=\sum_{a} f\left(\langle b|\hat{H}_{\rm
e-v}|a\rangle[E_a-E_b]^{-1}\right) w_a$ where
$f(x)=(1+x^{-2})^{-\frac 12}$,
the sum extends over all the states $|a\rangle$ in the previous tier,
and the weight of
$\hat{c}_{m \sigma}^\dagger | i \rangle$
is taken as unity.  }.
The application of about 350 Lanczos steps \cite{Prelovsek00}
generates a tridiagonal matrix, which provides a well converged
spectrum.

The basis for the calculation of Fig.~\ref{allmodesT800:fig}
includes about 720000 states, in particular all the states generated by up
to three applications of $\hat{H}_{\rm e-v}$ to $\hat{c}_{m \sigma}^\dagger
| i \rangle$: the convergence of the truncated basis is quite good.
We repeat the whole procedure of basis generation / spectrum calculation for
each of the $N_{\rm sample}$ initial states, and average the obtained
spectra.
The above procedure is carried out including all $H_g$ and $G_g$ modes. The
$A_g$ modes separate out, as they can be solved analytically as a simple
displaced oscillators, and are included exactly by convolution.

%
%

Figure~\ref{allmodesT800:fig} compares the theoretical PE 
spectrum calculated at $T=800$~K with the
measured spectra by two separate groups.
All experimental features, including the characteristically broad and
asymmetric main PE peak, the strongest satellite peak around 230~meV, a
second weaker peak near 400~meV, and a slow decay up to 600~meV are
remarkably reproduced, based purely on {\it ab initio} parameters.
The model and its ingredients thus appear to describe quite accurately the
PE spectrum of C$_{60}$. We can now take it apart, so as to understand the
underlying physics to satisfaction.

%
%

We address the two main features of the spectrum, namely a) the large
broadening and b) the 230~meV peak.
By eliminating $G_g$ and $A_g$ modes from the calculation there is no
large change in the spectrum. Thus only the $H_g$ modes are important.
Among $H_g$ modes, the largest couplings belong to the low-frequency
$H_g^{(1)}$ 32~meV quadrupolar radial mode, and to the two high-frequency
C-C stretch modes $H_g^{(7)}$ and $H_g^{(8)}$ at $\hbar\omega=181$ and
197~meV).
When we include $H_g^{(1)}$ alone we find (inset) that the thermally
broadened asymmetric main peak is well reproduced.
Therefore $H_g^{(1)}$ is the main responsible for the large
broadening.
%

When only $H_g^{(1)}$ is included, all satellite structures above 200~meV
are absent, and do not reappear even if we arbitrarily increase the
$H_g^{(1)}$ coupling to larger and larger values.
This rules out the possibility to attribute the 230~meV structure to an
``electronic'' splitting \cite{ManiniCommAndReply03}.
As it turns out, electronic JT splittings become visible in experimental PE
spectra, such as those of Fe(CO)$_5$ \cite{Hubbard81}, and in theoretical
spectra, like those of Martinelli {\it et al.}~\cite{Martinelli91etal} only
for very strong e-v coupling, whereas in C$_{60}$ coupling is intermediate
at most.
That suggests that the 230~meV peak in C$_{60}$ should rather be associated
to the high-frequency modes $H_g^{(7)}$ and $H_g^{(8)}$. However, the
connection between their frequencies $<200$~meV and the 230~meV
peak position is far from obvious.
To clarify this, we calculate the spectrum by including $H_g^{(7)}$ and
$H_g^{(8)}$ alone (inset of Fig.~\ref{allmodesT800:fig}).
The result is a much narrower spectrum ($H_g^{(1)}$ is now omitted),
showing a high-energy satellite at a significantly (10\%) higher energy
than $\hbar\omega_{H_g8}$.
This increase in vibronic excitation energy above the purely vibrational
$\hbar\omega$ for the $v$=1-phonon excitation of a trivial displaced
oscillator is in fact a characteristic signature of the dynamic JT effect,
and can be understood on the basis of symmetry.
For $H_g^{(7)}$ and $H_g^{(8)}$, $\hbar\omega$ is almost three times the
thermal energy: the main initial-state contribution to the spectrum comes
from $|i\rangle$ = the neutral molecule GS, of symmetry $A_g$.
The bare hole $\hat{c}_{m \sigma}^\dagger | i \rangle$ therefore carries
the orbital $h_u$ symmetry of the HOMO.
This state has nonzero overlap strictly only with final states $|f\rangle$
of the same symmetry $h_u$.
This {\em single symmetry channel}\ thus represents the only contributor to the
low-temperature PE spectrum according to \eqref{FermiGR}
\footnote{
This, incidentally, rules out the possibility to observe tunnel split states
\cite{Canton02etal} in PE, as they necessarily possess
{\em different} symmetries.
}.
The blue shift of the vibronic peak relative to the
bare vibration is now due to two general properties (in fact common
to all dynamical JT systems): (i) the coupling independence of the {\em
average} excitation energy in any vibronic multiplet at weak coupling (the
leading change is of order $g^4$) \cite{BersukerWeakCoupl}; (ii) the
increasing ``repulsion'' that the GS level exerts for increasing $g$ on all
excited states of the same symmetry, which pushes them upward.

\begin{figure}
\centerline{
\epsfig{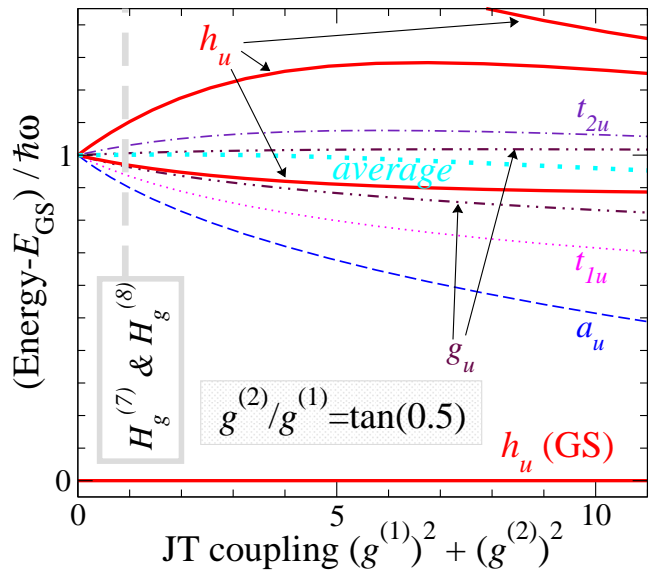}
}
\caption{\label{alpha0.5:fig}
Vibronic excitation energies for a single $H_g$ mode coupled to the $h_u$
level, as a function of coupling, for a ratio $g^{(2)}/g^{(1)}$
characteristic of the $H_g^{(7/8)}$ modes.
The {\it average} refers to the multiplet originating from the 1-phonon states.
Bold: $h_u$ states, the only visible states in $T=0$ PE.
}
\end{figure}

Both concepts are illustrated in Fig.~\ref{alpha0.5:fig}, depicting the
vibronic excitation energies (with special attention to the $h_u$ states)
for the simplest case of a single $H_g$ mode, as a function of the e-v
coupling strength $g^2$.
The approximate coupling independence of the {\em average} excitation
energy of the multiplet of states derived from the $v$=1-phonon manifold is
apparent.
Within this multiplet (made of $h_u\times H_g$ = $a_u + t_{1u} + t_{2u} + 2
g_u + 2 h_u$), the $h_u$ vibronic states are the sole that have another
state of the same symmetry (the GS itself) lower in energy ``pushing''
them upward.
As a result, for small but increasing coupling, the $h_u$ excited vibronic
states, move necessarily upward in energy above $\hbar\omega_{H_g}$.
Moreover the amount of shift is at weak coupling proportional to $g^2$
and reflects the strength of the coupling.
None of these results seems specific to the $h\otimes H$ JT model of
C$_{60}$. For example, identical conclusions apply to the $e\otimes
E$ problem \cite{BersukerWeakCoupl}.  Indeed a 15\% blue shift of the first
vibronic satellite is clearly observable in the PE spectra of benzene
\cite{Baltzer97etal}.

In C$_{60}$ we must consider the two high-frequency modes $H_g^{(7)}$
and $H_g^{(8)}$ at $\hbar\omega=181$ and 197~meV).
The couplings for these two modes alone accounts for a 10\%, or roughly
20~meV shift of the main $H_g^{(7,8)}$ vibronic satellite above
$\hbar\omega$.
This is about half of the total observed and calculated shift of $\sim
40$~meV of the 230~meV satellite relative to $\hbar\omega$.
The remainder is a collective effect of all other modes simultaneously
interacting with the HOMO, and pushing their respective $v=1$ satellites
(weak and thus invisible at large temperature) upward.
All goes as if effectively the coupling of the $H_g^{(7/8)}$ modes were
larger, i.e.\ effectively moving to the right of the vertical dashed line
in Fig.~\ref{alpha0.5:fig}.

\begin{figure}
\centerline{
\epsfig{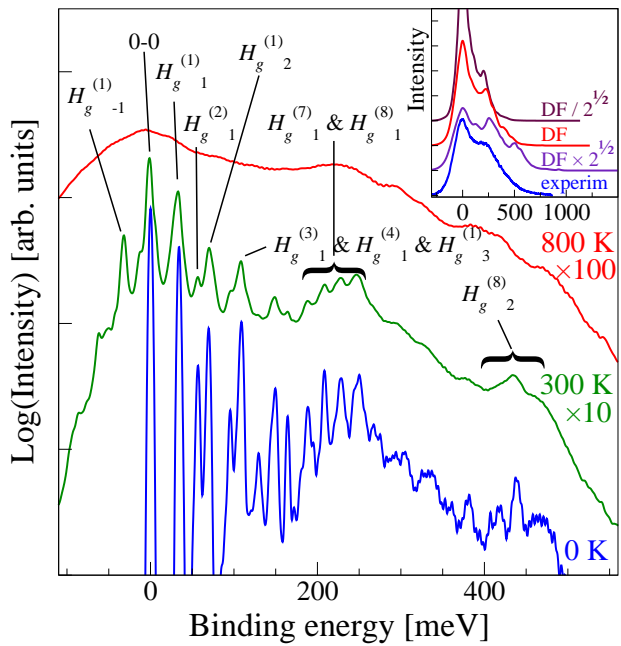}
}
\caption{\label{allmodesTdependence:fig}
Temperature dependence of the PE, computed including all modes of C$_{60}$
for $T=0$, $T=300$~K and $T=800$~K.
Labels ${\Lambda^{(j)}}_v$ indicate the leading component in the vibronic
states.
Inset: computed spectra ($T=800$~K) based on the same DF coupling parameters
$g_{\Lambda j}^r$ divided by $\sqrt 2$, original, and multiplied by
$\sqrt 2$, compared to experiment~\cite{Bruhwiler97etal}.
}
\end{figure}

We have thus attained the following understanding of the 800~K PE spectrum
of C$_{60}$: (i) the strong 230~meV satellite (and the weak 400~meV too) is
due to $h_u$-symmetry vibrons derived from $H_g^{(7/8)}$ modes, blue
shifted by repulsive coupling to the $h_u$ GS; (ii) the large
spectral broadening is due mainly to $H_g^{(1)}$.
Clearly, this broadening of the 800~K spectrum prevents direct extraction
of further detailed information about the actual e-v couplings of the
fullerene cation.
Our calculations, when repeated for lower temperatures, indicate a wealth
of structures that should emerge. 
Figure~\ref{allmodesTdependence:fig} shows that already a measurement
carried out at 300~K would provide enough detail to get precise indications
about the couplings of individual modes.
For example, the intensity ratios between the main peak and the subsequent
peaks at 35 and 65~meV is a direct measure of $g_{H_g1}$.
A cool molecular beam experiment \cite{Miller88} would be extremely useful.
%
Also in view of future 
cool-beam
data it is meaningful to address the question
of how accurate the calculated DF e-v couplings used here really are.
For $t_{1u}$ electrons coupled to $H_g$ vibrations, the calculated DF
couplings were argued for example to be too weak by a substantial amount,
by comparison to the larger couplings needed in order to fit C$_{60}^-$ PE
\cite{Gunnarsson95etal}.
%
%
In our HOMO case, the excellent agreement of calculations with experiment
suggests that there is no such failure.
Figure~\ref{allmodesTdependence:fig} (inset) shows several PE calculations
with rigidly rescaled e-v parameters. Comparison with experiment indicates
that the DF couplings are, on the whole, quite close (if slightly on the
weak side) to the actual couplings of C$_{60}^+$.
Much larger JT coupling values like those obtained using different approaches
\cite{Bendale92etal} are ruled out.
That is very good news, for it gives further confidence in the future
possibility to rely solidly on DF for a good description of the
intra-fullerene chemical bond, and of its perturbations.

%
%
%

In conclusion, we present here a detailed calculation of the PE spectrum of a
high-symmetry molecule, C$_{60}$, which is JT active in its one-hole state.
We find excellent agreement with experiment, confirming for the first
time the accuracy of the e-v couplings as extracted from DF calculations.
Perhaps more importantly, we reach a fresh understanding of the PE
spectrum, with its width, its vibronic peak, and the reason for its blue
shift relative to the bare vibrations.
The blue shift of the one-vibron satellite is argued to represent a more
general characteristic fingerprint of JT systems in the
weak-to-intermediate coupling regime, and is found for example in the
$e\otimes E$ JT of benzene as well.


We are indebted to G.P. Brivio, P.\ Br\"uhwiler, A.\ Del Monte, D.\ Galli,
M.\ Lueders, L.\ Molinari, G.\ Onida, F.\ Parmigiani, A.\ Parola, and G.E.\
Santoro, for useful discussion.
This work was supported by the European Union, contracts ERBFMRXCT970155
(TMR FULPROP) as well as by MIUR COFIN01, and FIRB RBAU017S8R
operated by INFM.




\end{document}